\def\kms{$\mathrm{km\;s}^{-1}$}

\def\h3{$h_{3}$}
\def\h4{$h_{4}$}

\def\grsim{{}^> _{\sim}}

\documentclass[structabstract]{aa}
\usepackage{graphicx}
\usepackage{txfonts}
\begin{document}
\title{MOND vs. Newtonian dynamics in early-type galaxies}

\subtitle{The case of NGC~4649 (M60)}

\author{S. Samurovi\'c
          \inst{1}
          \and
          M.M. \'Cirkovi\'c\inst{1}
      %\fnmsep
      %\thanks{}
          }

\institute{$^1$Astronomical Observatory, Volgina 7, 11160 Belgrade, Serbia\\
\email{srdjan@aob.bg.ac.yu}\\
\email{mcirkovic@aob.bg.ac.yu}}

   \date{Received 6 February 2008; accepted 12 June 2008}

% \abstract{}{}{}{}{}
% 5 {} token are mandatory

  \abstract
  % context heading (optional)
  % {} leave it empty if necessary
{Regarding the significant interest in both dark matter and the application of MOND to early-type galaxies, we investigate the MOND theory by comparing its predictions, for models of constant mass-to-light ratio, with observational data of the early-type galaxy NGC 4649.}
  % aims heading (mandatory)
{We study whether measurements for NGC 3379 and NGC 1399 are typical of early-type systems and we test the assumption of a Newtonian constant M/L ratio underlying most of the published models.}
  % methods heading (mandatory)
   {We employ the globular clusters of NGC~4649 as a mass tracer.
The Jeans equation is calculated for both MOND and constant
mass-to-light ratio assumptions. Spherical symmetry is assumed and the calculations are performed for both isotropic and anisotropic cases.}
  % results heading (mandatory)
   {We found that both Jeans models with the assumption of a constant mass-to-light ratio and  different MOND models provide good agreement with the observed values of the velocity dispersion.
   The most accurate fits of the velocity dispersion were obtained for the mass-to-light ratio in the $B$-band, which was equal to 7, implying that there is no need for significant amounts of dark matter in the outer parts
   (beyond 3 effective radii) of this galaxy. 
We also found that tangential anisotropies are most likely present in NGC~4649.}
  % conclusions heading (optional), leave it empty if necessary
   {}


   \maketitle
%
%________________________________________________________________

\section{Introduction}

The elliptical (early-type) galaxies are pressure-supported rather than rotationally supported systems and it is observationally more difficult to infer the presence of dark matter than in the more well-studied late-type galaxies.
This is because, at large galactocentric distances, ellipticals do not have a straightforward tracer of circular orbits that is similar to the neutral hydrogen used to study the halo kinematics of spirals.
Fortunately, however we are able to use several alternative tracers
in the nearby ellipticals: 1) integrated stellar spectra
(although the observations are limited practically within $\sim 3-4R_e$, where
$R_e$ is effective radius), 2) X-rays 3) planetary nebulae (PNe)
and 4) globular clusters (GCs).

Studies of ellipticals have indicated that within $\sim
2-3R_e$ dark matter does not appear to dominate (e.g.~Samurovi\'c \&
Danziger 2005, 2006). Beyond $\sim 3R_e$, dark matter appears to
start to play important dynamical role. Although the presence of dark matter
in the Universe is widely accepted, some alternative opinions have been expressed such as the theory of  MOND (Milgrom 1983;  recent
reviews in Scarpa 2006, Milgrom 2008). The case for giant
ellipticals has not been considered adequately and one
possible reason is the scarcity of available kinematical
data out to large galactocentric distances. Several attempts to model ellipticals with dark matter and/or MOND theory have
appeared (e.g.~ Schuberth et al. 2006, Klypin \& Prada 2007,
Tiret et al. 2007, Richtler et al. 2008) with ambiguous results.
Notably, Tiret et al. (2007) claimed that a MOND model for the
dynamics of NGC 3379 reproduced the observations on all scales, while
Richtler et al. (2008), studying NGC~1399, reached the opposite
conclusion -- that the best-fit MOND model still requires an
``additional hypothetical dark halo''. 
The latter result, for matter on small scales, agrees with observations implying the need for dark matter at the centers of clusters: for example, Angus, Famaey \& Buote (2007) found that a
hidden mass component (about 1.5 to 4 times more massive than the
total visible mass) is required to explain the hydrostatic
equilibrium of clusters with temperatures ranging from 0.7 to 8.9
keV.

In this paper, we discuss the early-type galaxy NGC~4649 (M60)
belonging to the Virgo cluster, for which new observational
data have become available (Lee et al. 2008a). This
galaxy is more typical than either a field giant elliptical
(NGC~3379) or the central galaxy of a cluster (NGC~1399), since it is a
typical cluster member. {\rm The plan of the paper is as follows:
in Sect.~2, we present the data related to NGC~4649;
in Sect.~3, we calculate the total mass of this galaxy for Newtonian
gravity and the MOND approach and solve the Jeans equation for the
constant mass-to-light ratio and MOND approach to determine
the best-fit parameters. Our conclusions are presented in
Sect.~4.}

\section{Observational data}

The  data used in this paper are based on observations of GCs in
NGC~4649 presented by Lee et al. (2008a). The
sample consists of 121 GCs (83 blue and 38 red  GCs). In all of our
calculations, we always consider the entire sample, i.e.~we do not
divide the data into subsets, so that we are able to have a larger number of GCs per bin in our analyses.

NGC~4649 (M60) is a giant elliptical galaxy in the Virgo cluster
that has a nearby companion, NGC~4647 (Sc galaxy at $2.'5$ from
the center of NGC~4649). The systemic velocity of NGC~4649 is
$v_{\rm vel}=1117\pm 6$ \kms. We consider  two values for the distance to NGC~4649: (i) the first one, $d=17.30$ Mpc  is based on
the surface brightness fluctuation method and is taken from Lee et al.~(2008a) (in this case, one arcsec corresponds to 84 pc); and (ii)
the second one is based on the aforementioned systemic velocity, which implies that using the WMAP estimate of the Hubble constant from Komatsu et al.~(2008) $h_0=0.70$, we obtain $d=15.96$ Mpc, for which one arcsec corresponds to 77.5 pc. Both values of the distance are tested below and the most appropriate fit is presented.
For the effective radius, we assume the value of $R_e=90$ arcsec (equal to $7.56$ kpc for $d=17.30$ Mpc, and $6.97$ kpc for $d=15.96$ Mpc),
which is the value taken from the paper by  Kim et al.~(2006). We note that
there are different estimates in the literature, for example, Lee
et al.~(2008a) assumed $R_e=110$ arcsec and according to the RC3
catalog (de Vaucouleurs et al.~1991), $R_e=69$ arcsec. The
discrepancy does not  impact significantly our present conclusions.

We used the radial velocities of GCs to determine the kinematics of NGC~4649: we
calculated the velocity dispersion, the skewness and kurtosis
parameters, $s_3$ and $s_4$, using standard definitions and the
NAG routine {\tt G01AAF}. We note that our intention is not
to reconstruct the full line-of-sight velocity distribution,
because it is known (see Merritt 1997) that for small data sets
such as the one we consider here, that contain less than a few hundred
objects per bin, this is impossible. We note, however, that
Wu \& Tremaine (2006) developed a maximum likelihood
method for determination the mass distribution in spherical
stellar systems that uses test particles and applied this method successfully
to a giant elliptical M87 using 161 GCs found in this
galaxy.

\begin{figure}
\centering
\includegraphics[width=6cm]{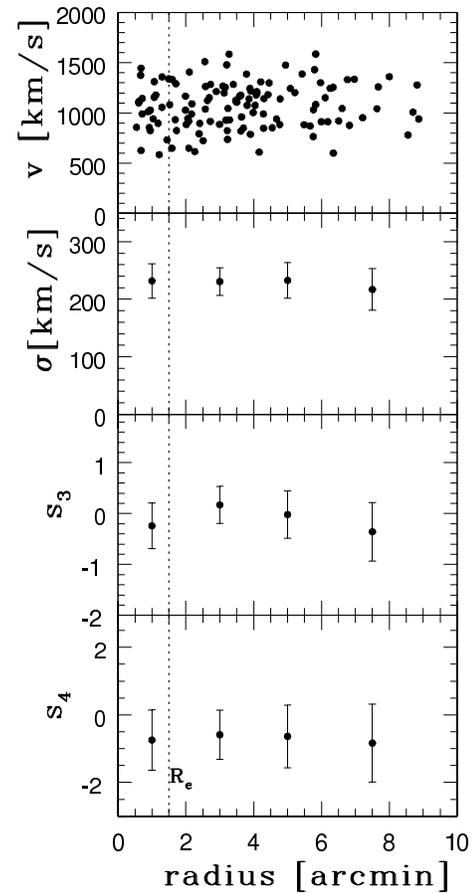}
\caption{Kinematics of   NGC~4649 based on the total sample of red and blue GCs. From top to bottom: radial velocity of the GCs in \kms; velocity dispersion calculated in a given bin; and the $s_3$ and $s_4$ parameters, which describe symmetric and asymmetric departures from the Gaussian, respectively.
}
\label{fig:kinematics_1}
\end{figure}

We therefore calculate skewness and kurtosis parameters
to determine whether, in some bins, a significant departure from a
Gaussian distribution (and  a tendency toward some
particular type of orbits) exists.  This is similar to the approach applied
by Teodorescu et al.~(2005) in their Fig.~18. Our results
regarding departures from a Gaussian distribution are applied below
 where we apply different anisotropies
when solving the Jeans equation. The results are given in
 Fig.~\ref{fig:kinematics_1} {\rm and in Table \ref{tab:table_1}}. What is obvious from the data (and
what makes this galaxy a good candidate for the dynamical
modeling based on two different techniques) is that the velocity
dispersion appears to be constant throughout the entire galaxy, and,
based on the values of $s_4$ (small negative values but
consistent with zero), there appears to be a tendency
towards tangential orbits; the same conclusion based on Jeans
modeling was reached by Hwang et al.~(2008)
(and also by Bridges et al.~(2006), who found that
the orbital distribution beyond 100 arcsec becomes tangentially
biased).

\begin{table*}%[ht!]
%\addtocounter{table}{+1}
\caption{Kinematics data for NGC~4649 for the total sample of clusters}
\begin{center}
\begin{tabular}{cccrcccc}
\hline
\noalign{\smallskip}
%\multicolumn{1}{c}{radial bin} &
\multicolumn{1}{c}{$<r>$} &
\multicolumn{1}{c}{$\sigma_{\rm total}$} &
\multicolumn{1}{c}{err$\_{\sigma_{\rm  total}}$}  &
\multicolumn{1}{c}{${s_3}_{\rm total}$} &
\multicolumn{1}{c}{err${\_{s_3}}_{\rm total}$} &
\multicolumn{1}{c}{${s_4}_{\rm total}$} &
\multicolumn{1}{c}{err${\_{s_4}}_{\rm total}$} &
\multicolumn{1}{c}{$N$} \\

%\multicolumn{1}{c}{  }\\
\noalign{\smallskip}
\multicolumn{1}{c}{(arcmin)} &
\multicolumn{1}{c}{(\kms)} &
\multicolumn{1}{c}{(\kms)} &
\multicolumn{1}{c}{ } &
\multicolumn{1}{c}{ } &
\multicolumn{1}{c}{} &
\multicolumn{1}{c}{} &
\multicolumn{1}{c}{} \\
\multicolumn{1}{c}{(1)} &
\multicolumn{1}{c}{(2)} &
\multicolumn{1}{c}{(3)} &
\multicolumn{1}{c}{(4)} &
\multicolumn{1}{c}{(5)} &
\multicolumn{1}{c}{(6)} &
\multicolumn{1}{c}{(7)} &
\multicolumn{1}{c}{(8)}  \\
\hline
\noalign{\smallskip}

1     &  232 &  30 &  -0.03   &    0.06&   -0.04& 0.04 &  30\\
3     &  231 &  24 &   0.02   &    0.05&   -0.03& 0.04 &  45\\
5     &  233 &  31 &   0.00   &    0.07&   -0.03& 0.05 &  28\\
7.5   &  217 &  36 &  -0.05   &    0.08&   -0.04& 0.06 &  18\\

\noalign{\smallskip}
\hline
\noalign{\medskip}
\end{tabular}
\end{center}
\begin{minipage}{18cm}
NOTES --
Col. (1):  central point for a given bin.
Col. (2):  velocity dispersion for the total sample of clusters.
Col. (3):  formal errors for the velocity dispersion of the total sample of clusters.
Col. (4):  $s_3$ parameter for the total sample of  clusters.
Col. (5):  formal errors for the $s_3$ parameter for the total sample of clusters.
Col. (6):  $s_4$ parameter for the total sample of clusters.
Col. (7):  formal errors for the $s_4$ parameter for the total sample of clusters.
Col. (8):  number of clusters in a given bin.

\end{minipage}
\label{tab:table_1}
\end{table*}

\begin{figure}
\centering
\includegraphics[width=8cm]{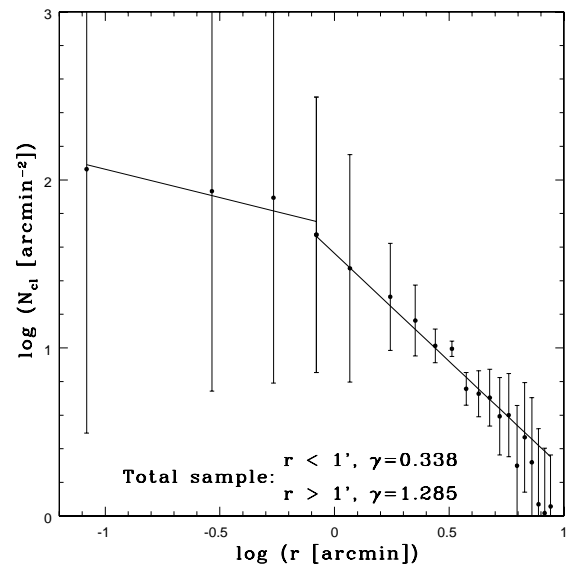}
\caption{Radial distribution of GCs (both blue and red) in NGC~4649. A power law is fitted to the radial surface density of GCs: $\Sigma\propto r^{-\gamma}$ (see text for details).}
\label{fig:surf_dens}
\end{figure}

For the Jeans modeling presented below, we also needed to determine
the radial surface density profile of the clusters in NGC~4649.
This plot is given in Fig.~\ref{fig:surf_dens} and represents
the radial distribution of the total sample of GCs in NGC~4649 as a
function of radius. The fit is given by a power law of the form:
$\Sigma\propto r^{-\gamma}$. Using the least squares method, we
decided that, within a galactocentric radius of approximately
1 arcmin $\gamma=0.338$, and for the region beyond 1 arcmin, that we should use the value $\gamma=1.285$ derived by
Lee et al.~(2008b).

\section{Models}
We model the observed velocity dispersion {\rm beyond 1 arc\-min (because our  observationally based points are all beyond $\sim 1$ arc\-min)} using the Jeans equation (e.g.~Binney \& Tremaine 1987):
\begin{equation}
{{\rm d}\sigma_r^2 \over {\rm d}r}
+ { \sigma_r^2 {(2\beta_*+\alpha)\over r}} = - {G M(r)\over r^2} + {v_{{\rm rot}}^2\over r}
\label{eqn:Jeans1}
\end{equation}
where  $\sigma_r$ is the radial stellar velocity
dispersion,\break $\alpha={\rm d}\ln \rho /{\rm d}\ln r$ is the slope of tracer density $\rho$ ({\rm the surface density is given above and in the models below we use $\alpha= -2.285$}).
The rotation speed $v_{{\rm rot}}$ was found to be non-zero, i.e.~ $v_{{\rm rot}}=141^{+50}_{-38}$ \kms\ (Hwang et al.~2008) and is used in all modeling below.
A parameter $\beta_*$ is introduced to describe the non-spherical nature of  the stellar velocity dispersion:
\begin{equation}
\beta_* = 1 - \frac{\overline{v_\theta^2}}{\sigma_r^2},
\label{eqn:B1}
\end{equation}
where $\overline{v_\theta^2}=\overline{v_\theta}^2+\sigma_\theta^2$.
For  $0 < \beta_* < 1$, the orbits are predominantly radial, and
 the line-of-sight velocity distribution is more strongly peaked than a Gaussian profile (positive $s_4$ parameter), and
for  $-\infty \le \beta_* < 0$ the orbits are mostly
tangential, so that the profile is more flat--topped than a Gaussian (negative $s_4$ parameter) (Gerhard 1993).

{In all models below, we calculated the  projected line-of-sight velocity dispersion  (e.g. Binney \& Mamon 1982; Mathews \& Brighenti 2003):
\begin{equation}
\sigma_p^2(R) = { \int_R^{r_t}  \sigma_r^2(r)
\left[ 1 - (R/r)^2 \beta_* \right]
\rho(r) (r^2 - R^2)^{-1/2} r {\rm d}r
\over
\int_R^{r_t} \rho(r) (r^2 - R^2)^{-1/2} r {\rm d}r }
\label{eqn:SIG1}
\end{equation}
where the truncation radius, $r_t$,  extends  beyond the observed kinematical point of highest galactocentric radius.
We assumed in all the cases that $r_t \sim 7R_e$. }

%ovde

\subsection{Newtonian mass-follows-light models}

A relation for determining the typical mass-to-light ratio in  elliptical
galaxies was given by van der Marel (1991), who found for
37 bright ellipticals that the average mass-to-light ratio in
the $B$--band in solar units was: $M/L_B=(5.93\pm 0.25)\, h_{50}$
becoming $M/L_B=(11.86\pm 0.50)\, h$ and therefore $M/L_B=8.30\pm
0.35$ for $h=0.7$. Since this study addressed the inner parts of
ellipticals, we considered the absolute upper limit for the
{\it visible (stellar)} component to be $M/L_B\sim 9-10$. An inferred
mass-to-light ratio of over approximately  $10$ in a given region would therefore imply the existence of dark matter.

For constant mass-to-light ratio models, we consider relations that
include stellar mass and dark matter distributed in the form of the standard
Hernquist (1990) profile:
\begin{equation}
\rho_H(r) = \frac{M_T}{2\pi} \frac{a}{r} \frac{1}{(r+a)^3},
\label{eqn:Hernquist}
\end{equation}
which has two parameters: the total mass $M_T$ and scale length,
$a$, where $R_e=1.8153a$. We solve the Jeans equation
(Eq.~\ref{eqn:Jeans1}) and consider values of $M/L_B$ of between 7 and
15, while repeating that $M/L_B \grsim 10$ is incompatible with an
entirely stellar component. As can be seen from Fig.~\ref{fig:NGC4649_cst},
we tested 3
values of the mass-to-light ratio  that are presented by the
hatched stripes: low ($M/L_B=7)$, intermediate ($M/L_B=10)$, and
high ($M/L_B=15$,  {\rm includes at least $\sim 50$\% of dark
matter}). In all cases, we allowed a realistic variation in both 
the $\beta_*$ parameter and the distance to the galaxy, $d$ for the fixed value of the mass-to-light ratio. If we need to identify the
constant mass-to-light ratio that most accurately
describes the observed velocity dispersion data, we would present the model with the following parameters: lower mass-to-light ratio, $M/L_B=7$,
slightly tangential orbits, $\beta_*=-0.2$, and the distance $d=17.30$ Mpc (see
Fig.~\ref{fig:NGC4649_cst} and Table \ref{tab:bestfit}).
We note that a good fit can also be achieved for a higher value of the mass-to-light ratio but then the $\beta_*$ parameter rises and the orbits become isotropic (see Fig.~\ref{fig:NGC4649_cst}, central ``S2'' stripe, which presents a case for which $M/L_B=10$).

With the assumption of the constant mass-to-light ratio the overall conclusion of the Jeans modeling is that, although we detect a hint of a rising mass-to-light ratio with increasing radius, due to the large error bars, we can obtain a satisfactory fit by also assuming a constant value $M/L_B\sim 7$, which implies that dark matter is not playing an important dynamical role, even beyond $\sim 3R_e$. The value of the constant mass-to-light ratio $M/L_B\sim 7$ implies that the total mass of NGC~4649 is equal to $\sim 5\times 10^{11}M_\odot$ at $\sim 3R_e$. This estimate is in agreement with predictions based on the   assumption that the Virial theorem holds (see Bertin et al.~2002; Cappellari et al.~2006).  This total mass is used below.

\subsection{MOND models}

We follow the recommendation of our referee and calculate the
total mass of NGC~4649 in MOND gravity using three different
formulas: (i) the ``simple'' MOND formula from Famaey \& Binney
(2005), (ii) the ``standard'' formula (Sanders \& McGaugh 2002),
and (iii) the Bekenstein's ``toy'' model (Bekenstein 2004). We
write the Newtonian acceleration as $a_N= a\mu(a/a_0)$, where $a$ is
the MOND acceleration, $\mu (x)$ is the MOND interpolating
function where $x\equiv a/a_0$, and  $a_0=1.35^{+0.28}_{-0.42}\times
10^{-8}$ ${\rm cm\,s^{-2}}$  is a universal constant (Famaey et
al.~2007). The interpolation function $\mu(a/a_0)$ shows an
asymptotic behavior, $\mu \approx 1$, for $a\gg a_0$, and we derive the Newtonian relation in the strong field regime, and
$\mu=a/a_0$ for $a\ll a_0$. The MOND dynamical mass, $M_M$, can be
expressed in terms of the Newtonian values, $M_N$ using the following
expression (e.g.~Angus et al.~2007):

\begin{equation}
M_M(r)= M_N(r) \times \mu(x).
\label{eqn:mass_mond}
\end{equation}
The intepolation function can have different forms as given below.
%%% simple mond
\begin{enumerate}
\item A ``simple'' MOND formula is given by:
\begin{equation}
\mu(x) = {x\over 1+x}.
\label{eqn:simple1}
\end{equation}

In this case, the circular velocity curve can be written as (e.g. Richtler et al.~2008)
\begin{equation}
V_{\rm circ,M}^2 = {V_{\rm circ,N}^2 \over 2}+\sqrt{{V^4_{\rm circ,N}\over 4}+V^2_{\rm circ,N}\times a_0\times r},
\label{eqn:vcirc_sim}
\end{equation}
where $V_{\rm circ,N}$ is the Newtonian circular velocity.
%%%% std mond
\item A ``standard'' MOND formula is given by:
\begin{equation}
\mu(x) = {x\over \sqrt{1+x^2}}.
\label{eqn:std1}
\end{equation}
It can be shown that the circular velocity curve then becomes:

\begin{equation}
V_{\rm circ,M}^4 = {V_{\rm circ,N}^4 \over 2}+\sqrt{{V^8_{\rm circ,N}\over 4}+V^4_{\rm circ,N}\times a_0^2\times r^2}.
\label{eqn:vcirc_std}
\end{equation}

%%% toy model

\item Finally, for the ``toy'' model the MOND formula is:

\begin{equation}
\mu(x) = {-1+\sqrt{1+4x}\over  {1+\sqrt{1+4x}}}.
\label{eqn:toy1}
\end{equation}

After some arithmetic, we derive the expression for the circular velocity curve:

\begin{equation}
V_{\rm circ,M}^2 = V_{\rm circ,N}^2 +\sqrt{a_0\times r} \; V_{\rm circ,N}.
\label{eqn:vcirc_toy}
\end{equation}
\end{enumerate}
These three MOND models are
tested below using Jeans models, and the best-fit model parameters are indicated in each case.
%}

{We  solved the Jeans equation (Eq.~\ref{eqn:Jeans1})
 based on the Hernquist profile (Eq.~\ref{eqn:Hernquist}), where $M_T$ is calculated using Eq.~\ref{eqn:mass_mond}.
We varied different parameters: distance $d$, parameter
$a_0$, and anisotropy $\beta_*$. The best-fit models are presented in
Fig.~\ref{fig:NGC4649_mond} and Table \ref{tab:bestfit}. We tested all three  MOND models
and it can be seen that the best fits were obtained for higher
value of the distance, $d=17.30$ Mpc. We note that the
quality of the fits for all MOND models was approximately  the same, as in the case of the
constant mass-to-light ratio models, as indicatated by the reduced $\chi^2$
values (see column 6 in Table \ref{tab:bestfit}). We formed our Jeans models
based on an expression for the mass corresponding to each of the constant mass-to-light ratios, as in the previous section. The lower mass-to-light ratio ($M/L_B=7$) provides a good fit to the observed velocity dispersion for all  tested models.
The preferred value of the constant $a_0=1.35\times 10^{-8}$ cm~s$^{-1}$
was found in two cases (for the ``simple'' and ``toy'' model), whereas a lower value was found for the ``standard'' model,  $a_0=0.93\times 10^{-8}$ cm~s$^{-1}$. In all tested models, we found that there is a detection of
tangential anisotropy,  the largest extent for the ``toy'' model ($\beta_*=-0.5$) and the smallest extent for the ``standard'' model ($\beta_*=-0.3$).
It is important to emphasize that we cannot exclude higher (but not extremely high, see below) mass-to-light ratio in our models; for example, the ``toy'' model with $M/L_B=9$, $a_0=0.93\times 10^{-8}$ cm~s$^{-1}$, $d=17.30$ Mpc, and ($\beta_*=-0.3$) also provides a satisfactory fit to the observed data with reduced $\chi^2= 2.90$.
In general, because of the large observational error bars at this stage, we cannot exclude the possibility  of an increase in the velocity dispersion in the outer parts, which would imply the increase of the total mass-to-light ratio and the existence of dark matter.
We emphasize that these models are based on a constant mass-to-light ratio and, therefore, a fixed higher value of the mass-to-light ratio, in principle corresponds to a higher velocity dispersion close to the center and a more realistic value in the outer parts of NGC~4649.

It is known that the ``external field effect'' is a
phenomenological requirement of MOND, which has strong
implications for non-isolated systems (Sanders \& McGaugh 2002),
such as NGC~4649. This effect is complicated and was analyzed
in detail by Wu et al.~(2007);  here, we provide only an
approximate estimate of this effect on our results. We follow Famaey et
al.~(2007b) and calculate the gravitational force per unit mass
exerted by the Virgo cluster on NGC~4649: \begin{equation} a_{\rm
ext}={(GM^b_{\rm Vir}a_0)^{1/2}\over d_c}\simeq 0.15a_0
\label{eqn:gext}
\end{equation}
where $M^b_{\rm Vir}=2.1\times 10^{13}M_\odot$ is the baryonic
mass of the Virgo inside $d_c\sim 1$ Mpc (which is the distance of
NGC~4649 from the center of the Virgo cluster; the distance to
NGC~4649 is equal to 17.30 Mpc and is also used) taken from the paper by
McLaughlin (1999)  (his figure 2) who estimated that a global
baryon fraction in Virgo is $\sim 7\% $. We then apply the
external effect by inserting $\mu(|{\bf a}+{\bf a}_{\rm
ext}|/a_0)$ only in the case of the ``toy'' model to infer the
effects on the Jeans modeling. The result is plotted in
Fig.~\ref{fig:NGC4649_mond} with the dotted line (and is also
given in Table \ref{tab:bestfit}). The modeled velocity
dispersion is higher than in the case when we neglect the
external field but not by much. This result is again consistent with
a hypothesis of no dark matter. Parenthetically, the same reasoning
applies much more forcefully to the acceleration due to the
presence of the galaxy companion NGC~4647, for which the value in
Eq.~\ref{eqn:gext} is smaller for between one and two orders of
magnitude, depending on the baryonic mass estimate for the
companion.

The same conclusion regarding the dynamical importance of dark
matter reached in the case of the constant mass-to-light ratio
also holds for the Jeans models based on MOND: successful fits
were obtained without dark matter throughout the entire galaxy for
all MOND approaches. 
%} 
\begin{figure}
\centering
\includegraphics[width=8cm]{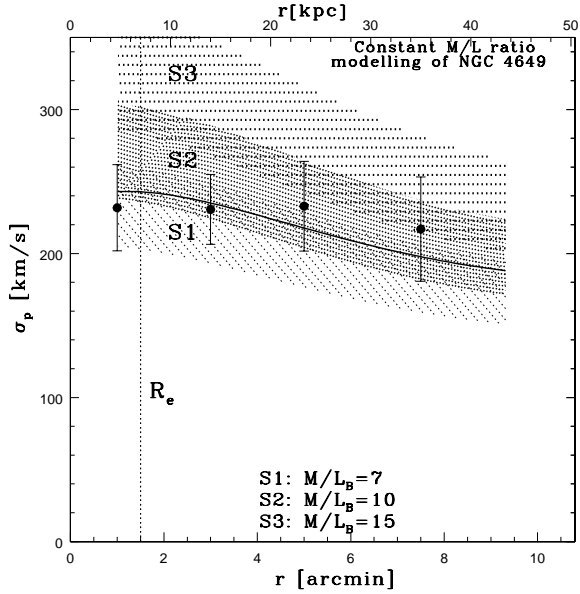}
\caption{\rm Jeans modeling of NGC~4649 using the constant M/L
approach. Three different stripes, which correspond to three
different constant mass-to-light ratios, are given and the upper
limit is always determined for $\beta_*=-0.2$, and $d=15.96$ Mpc while
the lower limit is determined for $\beta_*=0.2$ and  $d=17.30$
Mpc. The lowest stripe ``S1'' is for the constant mass-to-light
ratio $M/L_B=7$, the stripe in between, ``S2'' is for $M/L_B=10$,
and the upper stripe, ``S3'' is for $M/L_B=15$. The solid line is
the best fit obtained for the constant mass-to-light ratio
models: in this case $M/L_B=7$, $\beta_*=-0.2$ and $d=17.30$ Mpc.
The upper scale corresponds to the distance $d=17.30$ Mpc. }
\label{fig:NGC4649_cst}
\end{figure}

\begin{figure}
\centering
\includegraphics[width=8cm]{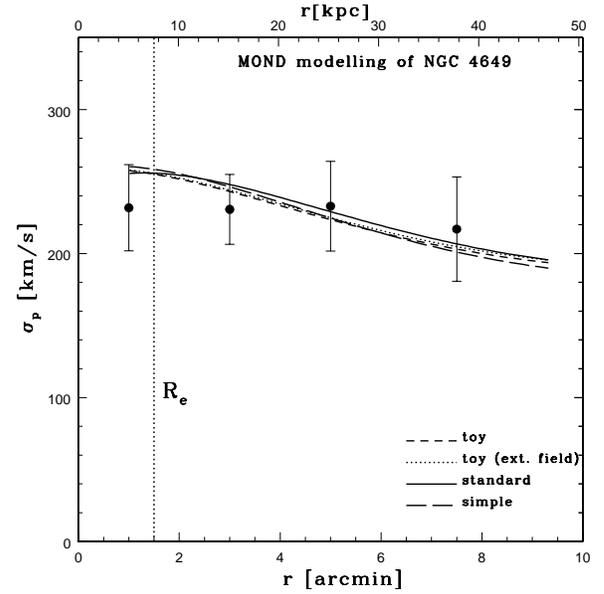}
\caption{\rm Jeans modeling of NGC~4649 using the MOND approach
and $M/L_B=7$.
Solid line is for the ``standard'' model for which $a_0=0.93\times 10^{-8}$ ${\rm cm \;
s^{-2}}$ and $\beta_*=-0.3$. Long-dashed line is for the
``simple'' model  for
which $a_0=1.35\times 10^{-8}$ ${\rm cm\; s^{-2}}$ and
$\beta_*=-0.4$. For the ``toy'' model we present two fits
for which $a_0=1.35\times 10^{-8}$ ${\rm cm\; s^{-2}}$ and $\beta_*=-0.5$: (i)
short-dashed line for the case when the external field is absent (ii) dotted line for the case when the external field is present. In this figure the upper scale corresponds
to $d=17.30$ Mpc because the best fits in all the
cases were obtained using this distance. } \label{fig:NGC4649_mond}
\end{figure}

\begin{table*}%[ht!]
%\addtocounter{table}{+1}
\caption{Best fits for constant M/L ratio and MOND modeling}
\begin{center}
\begin{tabular}{lcccrc}
\hline
\noalign{\smallskip}
%\multicolumn{1}{c}{radial bin} &
\multicolumn{1}{c}{Model} &
\multicolumn{1}{c}{$M/L_B$} &
\multicolumn{1}{c}{$a_0$ [$10^{-8}$ ${\rm cm \; s^{-2}}$]} &
\multicolumn{1}{c}{$d$ [Mpc]}  &
\multicolumn{1}{c}{$\beta_*$} &
\multicolumn{1}{c}{$\chi^2$} \\
\multicolumn{1}{c}{(1)} &
\multicolumn{1}{c}{(2)} &
\multicolumn{1}{c}{(3)} &
\multicolumn{1}{c}{(4)} &
\multicolumn{1}{c}{(5)} &
\multicolumn{1}{c}{(6)} \\
\hline
\noalign{\smallskip}

Const                    & 7 &  ---  &  17.30 & -0.2    &    1.36  \\
MOND Simple              & 7 &  1.35 &  17.30 & -0.4    &    2.31  \\
MOND Standard            & 7 &  0.93 &  17.30 & -0.3    &    1.66  \\
MOND Toy                 & 7 &  1.35 &  17.30 & -0.5    &    1.83  \\
MOND Toy (ext. field)    & 7 &  1.35 &  17.30 & -0.5    &    1.78  \\

\noalign{\smallskip}
\hline
\noalign{\medskip}
\end{tabular}
\end{center}
\begin{minipage}{18cm}
NOTES -- Col. (1):  the name of the model; ``ext. field'' is for the
MOND toy model for which the external field was taken into account (see text for details).
Col. (2): mass-to-light ratio in the $B$-band used.
Col. (3):  the
best-fit function value of the constant $a_0$; it is not used in the
case of the model with the constant mass-to-light ratio. Col.
(4):  the best fitting value of distance to NGC~4649. Col. (5):
the best fitting value of the $\beta_*$ parameter. Col. (6):
reduced $\chi^2$ of the best fit for a given model.

\end{minipage}
\label{tab:bestfit}
\end{table*}
\section{Conclusions}
The GCs kinematics of NGC~4649 was studied out to 260 arcsec
($=2.9R_e$ using the value of effective radius adopted in this
paper) by Bridges et al.~(2006). They found, using
isotropic and axisymmetric orbit-based models, that dark matter
exists in the halo of this galaxy.

In this paper, we have studied the dynamics of the early-type galaxy NGC~4649 (M60) for both Newtonian and MOND approaches using the Jeans equation and found that there is no need for significant amount of dark matter in its outer parts.
The principle results of our study were: 

\begin{enumerate}
\item We studied the kinematics of NGC~4649 and found that the velocity dispersion remains approximately constant throughout the entire galaxy. We detected an observational
 hint of small tangential anisotropies in this galaxy, which agrees with other results in the literature (Bridges et al.~2006, Hwang et al.~2008).

\item {We used the Jeans equation to infer the mass-to-light ratios that
can describe the observed velocity dispersion profile of NGC~4649: we found that
the successful fit to the velocity dispersion can be obtained using
$M/L_B\sim 7$ with moderate tangential anisotropies that indicate the lack of dark matter throughout this galaxy.
Higher values of the mass-to-light ratios ($M/L_B\sim 9-10$), which are still consistent with a no dark matter hypothesis, are also permitted but then the anisotropies tend to vanish.}

\item {We also used the Jeans equation to test the predictions of the MOND theory using three different models: ``simple'', ``standard'', and ''toy''.
We found that all the tested MOND models  provided a good fit to the observational velocity dispersion data (again for $M/L_B\sim 7$ with moderate  tangential anisotropies).

\item We have also briefly addressed the problem of the surrounding environment, since NGC~4649 belongs to the Virgo cluster.
We found that the external field at the position of NGC~4649 does not significantly alter our conclusions: the Bekenstein ``toy'' model again provides a good fit to the observed velocity dispersion (assuming $M/L_B\sim 7$) and is consistent with a no dark matter hypothesis.}
\end{enumerate}
 
\begin{acknowledgements}
We thank H.~S.~Hwang for providing us the data in electronic
form. This work was supported by the Ministry of Science of the
Republic of Serbia through the project no.~146012, ``Gaseous and
stellar component of galaxies: interaction and evolution''. This
research has made use of the NASA/IPAC Extragalactic Database
(NED) which is operated by the Jet Propulsion Laboratory,
California Institute of Technology, under contract with the
National Aeronautics and Space Administration.  We acknowledge
the usage of the HyperLeda database (http://leda.univ-lyon1.fr).
We thank the anonymous referee for the very useful and detailed
comments which helped to significantly improve the quality of the
manuscript.
\end{acknowledgements}

\end{document}